# How Does Artificial Intelligence Improve Human Decision-Making? Evidence from the AI-Powered Go Program[*]

Sukwoong Choi[†] Hyo Kang[‡] Namil Kim[§] Junsik Kim[**]

January 2025

***Abstract.*** We study how humans learn from AI, leveraging an introduction of an AI-powered Go program (APG) that unexpectedly outperformed the best professional player. We compare the move quality of professional players to APG's superior solutions around its public release. Our analysis of 749,190 moves demonstrates significant improvements in players' move quality, especially in the early stages of the game where uncertainty is highest. This improvement was accompanied by a higher alignment with AI's suggestions and a decreased number and magnitude of errors. Young players show greater improvement, suggesting potential inequality in learning from AI. Further, while players of all skill levels benefit, less skilled players gain higher marginal benefits. These findings have implications for managers seeking to adopt and utilize AI in their organizations.

***Managerial Abstract.*** We examine how professionals can learn from AI by studying an AI-powered Go program (APG) that outperformed the best professional player. By analyzing 749,190 moves, we find that players' move quality improved significantly, closely aligning with the AI's recommendations. The number and magnitude of errors also decreased. This learning effect was particularly strong early in the game where decisions are more uncertain. Young players showed greater effect, suggesting that learning from AI may vary by age. While players of all skill levels benefited, those with less skill saw the greatest improvement. These findings highlight the instructional role of AI and offer guidance on how to effectively integrate AI into organizations to enhance worker performance across different age groups and skill levels.

---

[*] We appreciate the insightful comments from the Associate Editor Evan Starr and two anonymous reviewers. We gratefully acknowledge valuable comments from Sinan Aral, Paola Criscuolo, Fabian Gaessler, Alfonso Gambardella, Nan Jia, Thorbjørn Knudsen, Myriam Mariani, Amalia R. Miller, Abhishek Nagaraj, Henning Piezunka, Ananya Sen, Neil Thompson, Christopher Tucci, Dashun Wang, and Yian Yin. We also thank organizers and audiences of the 2021 NBER Economics of AI Conference, the Fall 2021 MIT IDE Lunch Seminar, the 2021 Academy of Management Meeting, the 2021 Strategic Management Society Annual Conference, the 2021 Conference on Artificial Intelligence, Machine Learning, and Business Analytics, the 2022 IP and Innovation Seminar, the 3rd AI and Strategy Consortium, the 2022 Wharton Technology and Innovation Conference, the 2022 Columbia/Wharton Management, Analytics and Data Conference, and seminar participants at Bocconi University, Imperial College London, MIT, Northwestern University, Rotterdam School of Management, Stevens of Institute Technology, Stony Brook University, and University at Albany. We extend our gratitude to Hong-Ryeol Lee and Minho Kwak for providing proprietary data on Go playing time and their participation in interviews. This study was supported by the Institute of Management Research at Seoul National University and was previously circulated under the title, "Strategic Choices with Artificial Intelligence." All errors are our own.
[†] Massry School of Business, University at Albany, State University of New York. schoi27@albany.edu.
[‡] SNU Business School, Seoul National University. hyokang@snu.ac.kr.
[§] Corresponding author. Konkuk University. namilkim@konkuk.ac.kr.
[**] School of Engineering and Applied Sciences, Harvard University. mibastro@gmail.com.

## 1. INTRODUCTION

Artificial intelligence (AI) has developed substantially to date, and its capabilities have reached or even surpassed those of humans in numerous domains (Rai et al., 2019).[1] For instance, AI has outperformed human experts in strategic gameplay (Silver et al., 2017), medical diagnosis (Kim et al., 2021), bioinformatics (Senior et al., 2020), and drug discovery and development (Savage, 2021; Smalley, 2017). The rapid advancement of AI is transforming the future of professional work (De Cremer, 2020). In particular, AI helps workers perform better because it provides real-time assistance with their tasks (e.g., Allen & Choudhury, 2022; Choudhury et al., 2020; Lebovitz et al., 2022; Tong et al., 2021). Consultants utilizing AI completed more tasks and produced higher-quality work than their counterparts who did not use AI (Dell'Acqua et al., 2023). By comparing their own judgments to those provided by AI-based solutions, medical professionals improve the quality of their diagnoses (Lebovitz et al., 2022). By suggesting standardized medical codes, AI improves the productivity of coders and the quality of charts (Wang et al., 2019).

To understand the role of AI in workplaces, studies have examined its potential to assist humans through better division of labor (Dell'Acqua et al., 2023; Lebovitz et al., 2022) or to substitute for human workers (Acemoglu et al., 2022; Eloundou et al., 2023). However, other contexts—those where labor tasks cannot be divided with AI or where AI cannot replace humans—have received relatively little attention; this is because of challenges in assigning responsibility for outcomes to AI and because of morality, security, or privacy concerns related to AI. Examples of these challenges include handling classified information, making medical-ethics decisions in areas such as organ transplantation, and judicial sentencing. Further, AI's decisions are often met with resistance from those affected by them, and AI may lack the emotional intelligence needed to effectively deliver its outputs, especially in fields like counseling and education. In such scenarios, an optimal approach may involve human workers learning from AI and improving their intrinsic skills to better perform tasks themselves.

[1] We focus on the technical definition of AI as *deep-reinforcement-learning algorithms*, as detailed in Plaat (2022).

This study examines whether and how human professionals learn from AI, along with the mechanisms by which their performance is improved. We also shed light on AI's differential effects (Allen & Choudhury, 2022; Choudhury et al., 2020) by considering individuals' openness to new technologies and their ability to utilize these technologies (Barth et al., 2020; Tams et al., 2014). Specifically, we examine the ages of individuals and their skill levels as key factors that could affect AI's instructional impact.

Studying this topic empirically can be challenging due to several difficulties: finding a context where AI can train human professionals (but does not perform the task directly); observing a decision (or a series of decisions) by humans and assessing the results; and disentangling AI's clout on such decisions. Further, given that AI's dramatic progress is only recent, the limited availability of data has constrained researchers from examining its impact (Seamans & Raj, 2018).

We study professional players of Go, a strategy board game that provides a unique opportunity to overcome these challenges. Over thousands of years, professional Go players have accumulated knowledge, experience, and skill in this game. Yet the introduction of an AI-powered Go program (henceforth, APG) suddenly changed the way Go players learn and play the game. In the historic Go match (AlphaGo vs. Lee Sedol) held in 2016, AI beat the best human professional player for the first time and by a large margin. Shortly after this event, in 2017, the first open-access APG, Leela, which was far superior to the best professional player, became available to the public. Since Leela's release, professional Go players have used APGs heavily in their learning.

The great advantage of the Go context is that it allows us to observe every single decision of each player; the entire move history is well archived and maintained for all major games occurring in the last few decades. Furthermore, an APG can calculate the probability of winning for every move and can even perform these calculations for earlier games, those played before the APG was released. We calculated the winning probability of 749,190 moves by 1,241 professional Go players in 24,973 major games held from 2015 through 2019. We then assessed the quality of *each move* by comparing the probability of winning associated with the focal move to that of the APG's best solution. Note that professional Go players are not allowed to use APGs (or any other tools) in a professional match. Thus, any changes in move quality (or in the probability of winning) after an APG release are attributable to changes in human capabilities (i.e., learning). Even without real-time assistance from an APG, players increasingly made moves that coincided with the APG's best solutions for a given situation, demonstrating a genuine learning effect from AI.

A recent work, Gaessler and Piezunka (2023) (henceforth, GP), studied the role of AI as a training partner for chess players. By leveraging the staggered access to chess computers for Western European players (from 1977) and Soviet players (from 1989), GP finds that players with initially inferior skills and limited training opportunities could benefit from training with chess computers. Such AI-backed simulations provided players with ample training opportunities, addressing the scarcity of available human counterparts. Chess computers at that time, however, were limited to serving as imperfect training partners (but not necessarily as instructors) because their performance, despite rapid improvement, had not yet reached the level of the best players.

Building upon GP's inspiring work, we look beyond the training role of AI and to consider its *instructional* role in a more advanced and contemporary context. The Go context of the late 2010s is suitable for studying the full capabilities and impact of modern AI technologies, which are driven by machine learning and deep learning. AlphaGo in 2016 possessed computing power exceeding 7,500 times that of Deep Blue, the most advanced chess computer prior to 2000; in terms of floating point operations per second (FLOPS), AlphaGo surpasses Deep Blue by more than 80,000 times (Thompson et al., 2022). This superior computing power, along with high accuracy in outcome predictions (further demonstrated by the defeat of the best human player) suggests that APGs plays a key instructional role from which players can learn and refine their decision-making capabilities.

We find that, similar to the impact of chess computers on players, AI has improved human performance. Before APGs, the winning probability of each move by professional Go players averaged 2.47 percentage points lower than the best solution suggested by the APG. This gap decreased by about 0.756 percentage points on average (or 30.5%) and up to 1.3 percentage points (or 47.6%) after APG's public release. Importantly, the improvement was accompanied by an increased *alignment* between players' moves and AI's top suggestions, confirming that the effects were indeed driven by learning from AI. However, unlike GP, we show that learning from AI leads to improvements across all skill levels; it does not benefit only lower-skilled or less-trained players as they observed. This is most likely attributable to the advancements and broad accessibility of modern AI, which has far surpassed human capabilities. Furthermore, we find the improvement to be more prominent among younger players who are open to and capable of utilizing APGs.

We also explore the mechanisms through which professional players achieve a higher probability of winning. Our mediation analysis reveals that a focal player's improvement in move quality is achieved mainly by reducing *Errors* (the number of moves where the winning probability drops by 10 or more percentage points compared to the winning probability of the immediately preceding move) and by reducing *Critical mistake* (the magnitude of the biggest drop in winning probability during the game). Additional analyses indicate that the improvement in move quality eventually leads to winning the game. This effect is most prominent in the early stage of a game where uncertainty is higher and there is more opportunity for players to learn from AI.

This study, along with that of GP, is among the first to provide micro-level evidence of the instructional role of AI in human decisions and performance. Our empirical analysis of 749,190 moves in Go games has meaningful implications for AI's instructional role, notably for how it could educate and nurture professional decision-making capabilities in fast-paced, uncertain environments. Further, the fact that the young benefit more from APGs has important implications for digital literacy and for potential inequality in accessing, adopting, and utilizing AI. Finally, our findings highlight the boundary conditions and heterogeneity of AI's effectiveness (e.g., by age and skill level of workers) of which managers should be aware for successful adoption and utilization of AI in organizations.

## 2. AI AND DECISION-MAKING

### 2.1. The impact of AI on human decision-making

When making decisions, humans tend to draw on their conceptualization of the future as input into the decision-making process. Humans also depend on knowledge of causality, which they actively develop to understand how past actions impact future outcomes. Through these processes, humans can judge and learn from situations—even unexpected situations—to improve their decision-making processes and outcomes (Lindebaum et al., 2020; Mintzberg, 1994). However, individuals are limited in their ability to process information, which slows learning and limits its scope (Galbraith, 1974; Simon, 1955). This in turns leads to failure to optimize decision-making (Kalberg, 1980). To mitigate these biases and errors, researchers propose to set goals and aspirations to guide decision-making and to use backward- and forward-looking decision models (Chen, 2008; Gavetti & Levinthal, 2000). However, the benefits of these choice models are marginal in alleviating the aforementioned limitations to optimal decision-making.

Literature on information technology (IT) provides yet another set of solutions and argues that the adoption and utilization of new technologies compensate for these shortcomings. Information theory (e.g., Blackwell, 1953) and the information-processing view of the organization (Galbraith, 1974) propose that the more accurate and precise the information used in decision-making, the higher the firm's performance. This is primarily because IT improves a firm's ability to collect, analyze, and process information for internal operational decisions. Specifically, IT complements organizational practices, which in turn leads to higher productivity (Bapna et al., 2013; Brynjolfsson & Hitt, 2000). The positive relationship between the volume and quality of information and optimal decision-making has been supported by a plethora of studies (e.g., Brynjolfsson et al., 2011; Davenport & Harris, 2017).

As data availability has grown, researchers have extended these arguments to data-driven decision-making. The data about consumers, suppliers, competitors, partners, and the utilization of large-scale analytics have supported decision-making (Brynjolfsson et al., 2011). For example, Brynjolfsson et al. (2011) found that the adoption of data-driven decision-making practices increases financial returns. Saunders and Tambe (2013) revealed that firms with data-driven decision-making at an executive level have higher productivity and higher market valuations. Data analytics also support decision-making for R&D search and for incremental process improvements (Wu et al., 2020). Overall, the adoption and utilization of new IT plays an important role in decision-making at both organizational and individual levels.

Researchers have recently extended this discussion to the adoption and utilization of AI. The advance in AI with the development of machine learning and deep-learning algorithms contributes to the avoidance of mistakes and errors stemming from human judgments (Danziger et al., 2011). AI algorithms are fundamentally different from traditional data-driven approaches for several reasons (Agrawal et al., 2018). First, AI can make inferences by self-learning. AI, therefore, is better suited for discovering hidden patterns and can conduct insightful tasks that need human-like "intuition." Second, AI performs predictions and judgments with high accuracy, and the accuracy increases exponentially with the number of training sessions and the quantity of data. With AI, therefore, humans can revisit their decision-making practices that may otherwise have yielded inferior decisions. Thanks to a predictive capability, which is superior to that of classical statistics and econometric techniques, AI algorithms have been applied to a variety of different decision-making problems (Athey & Imbens, 2019).

These distinct characteristics enable AI to outperform humans not only in repetitive work and recognition tasks but also in creative tasks in some domains (He et al., 2015; Mnih et al., 2015). Researchers find that AI performs well even in high-level cognitive tasks such as making a legal decision in court (Kleinberg et al., 2017), discovering protein structure in biology (Senior et al. 2020), and playing strategic games (Schrittwieser et al., 2020), among other settings. Considering the assumption of bounded rationality—that decision-makers tend to balance the quality of their decisions with the cost, such as the cognitive effort and time required to reach their decisions (Kahneman, 2003)—AI can contribute to lowering cost, which in turn rebalances the accuracy of decisions. In other words, AI helps human decision-making by evaluating a broader scope of options at a lower cost and by performing a more accurate evaluation of the options available. For example, when a radiologist uses AI to read a chest X-ray, within a few seconds AI can show the probability of the patient having some predefined disease. Similarly, when professional Go players use AI, they can immediately learn the winning probability associated with each possible move and can distinguish better moves.

Based on AI's superior predictive power, managers have several incentives to learn from AI. First, classical decision-making theory proposes three conditions that face humans making decisions: certainty, risk, and uncertainty (Langholtz et al. 1993). Without knowing values associated with each choice, individuals make decisions under uncertainty, which may lead to unfavorable outcomes. AI, in contrast, provides accurate, predicted values and thereby reduces the uncertainty associated with choices. Managers who learn from AI therefore can make decisions under less uncertainty.

Second, the unified theory of acceptance and use of technology (Venkatesh et al., 2003) emphasizes that managers actively accept and utilize IT when they expect superior performance from its use. Informed managers should thus actively adopt AI in decision-making processes and consequently will achieve superior performance.

Lastly, managers who utilize AI learn to improve their decision-making ability. AI does not yet explain why a particular choice has better outcomes (Hagendorff & Wezel, 2020), but it can provide feedback on whether an individual choice is good or poor. By repeatedly comparing their choices with those of AI, managers can update or revise their evaluation criteria based on AI's feedback (Yechiam & Busemeyer, 2005). For instance, AI-powered simulations present managers with opportunities for experiential learning, enabling them to understand superior choices through direct experience (Gaessler & Piezunka, 2023). Therefore, being equipped with the ability to make better evaluative choices, managers can make better decisions even without real-time AI assistance.

**2.2. Differential adoption and utilization of AI by age**

AI has strong potential to train employees and improve their decision-making, but not all professionals benefit from AI to the same extent. Despite its superior performance in prediction, AI and its related products and services are relatively new and do not have a proven record in terms of credibility and stability. Professionals thus perceive AI-powered tools as generally riskier to adopt or utilize when making important decisions, with a tradeoff between performance and risk (Cadario et al., 2021; Lebovitz et al., 2021). The literature on the differential effects of AI highlights the role of an individual's age in the use of digital technology (Barth et al., 2020). Prior studies suggest that age is an important factor in adopting and utilizing new technology (Weinberg, 2004). Notably, the learning-by-doing literature indicates that the marginal effect of learning from new technology varies with age (or tenure) (Allen & Choudhury, 2022; Foster & Rosenzweig,

1995).

In the context of AI, extant studies have found mixed results. Wang et al. (2019) studied medical coders in hospitals who used AI suggestions for chart coding and found that the productivity of younger employees improved more than that of older employees. In contrast, Choudhury et al. (2020) found that senior employees, who possessed greater domain expertise than younger workers, tended to gain more complementary benefits from AI. Allen and Choudhury (2022) then suggest an inverted U-shaped relationship wherein employees with moderate experience are better able to utilize the algorithm tool. These studies tend to assume that seniority is associated with workers who have accumulated experience and breadth of knowledge. To better understand the differential effects of age-related learning on AI adoption and utilization, we draw on the literature examining algorithmic aversion and vintage-specific human capital.

***Algorithmic aversion***. Algorithmic aversion is the tendency of individuals to distrust or avoid algorithms in decision-making (Dietvorst et al., 2015). Individuals tend to undervalue the performance of algorithms, even when presented with evidence of the algorithm's superiority (Logg et al., 2019). Prior studies suggest that this aversion is exacerbated when individuals exhibit a higher level of risk aversion (Kahneman et al., 2016; Kahneman & Tversky, 1979), find a lack of transparency in the algorithm's workings (Shin & Park, 2019), and demonstrate low familiarity with technology (Dietvorst et al., 2015).

The degree of algorithmic aversion can also vary depending on the ages of individuals (Mahmud et al., 2022). Young professionals tend to exhibit lower levels of risk aversion than their older counterparts (Tyler & Steensma, 1998). Furthermore, older professionals tend to view algorithmic decisions as less beneficial (Araujo et al., 2020) and to exhibit lower trust in them (Lourenço et al., 2020). Allen and Choudhury (2022) show that senior professionals are more reluctant to accept algorithmic advice than are their younger colleagues, because seniors have greater confidence in their own expertise and a greater sense of accountability for their actions. Building on these insights, we argue that younger professionals are less prone to algorithmic aversion.

***Vintage-specific human capital***. Vintage-specific human capital refers to a unique set of skills and knowledge that is specific to a certain time period or technology (Chari & Hopenhayn, 1991). As technology evolves and tasks change, individuals with vintage-specific human capital are better equipped to adapt to and utilize new technologies effectively (Autor et al., 2003; Gibbons & Waldman, 2004). Younger professionals, having grown up in a digital environment from an early age, typically possess a better understanding of new technologies than do their elders. The learning-by-doing literature suggests that these experiences improve their knowledge and skills (Arrow, 1962; Foster & Rosenzweig, 1995) and equip them with rich vintage-specific human capital related to emerging technologies (Morris & Venkatesh, 2000; Schleife, 2006). Hence, younger professionals have greater absorptive capacity for AI intricacies and are more likely to learn effectively from AI.

### 2.3 Research questions

Based on the arguments above, we ask two primary questions about the relationship between AI and human decision-making. First, does AI improve the decision-making of human experts and, if so, how? Second, how does the influence of AI vary among human professionals according to their age and other characteristics? We argue that young professionals have greater incentives and greater ability to utilize AI-powered tools and to benefit from them. In what follows, we empirically examine these research questions and conduct a series of post-hoc analyses to unpack what drives the observed patterns.

## 3. EMPIRICAL STRATEGY

### 3.1. Setting

***The game of Go and professional tournaments.*** Go (or Baduk) is a two-player strategy board game that originated in China at least 3,000 years ago. The board consists of a grid of nineteen lines by nineteen lines. Players compete to control more of the board's territory, with one player using black stones and moving first, while the other uses white stones and moves second, alternating the placement of stones at the intersections of the lines. The professional Go industry is substantial—especially in China, Japan, South Korea, and Taiwan. More than ten major professional tournaments, sponsored by large corporations, are held throughout the year in each country. For example, the Kisei tournament in Japan—held annually since 1977 and sponsored by the *Yomiuri Shimbun* newspaper—awards 4,500,000 yen ($413,000) to the first-place winner in addition to per-game compensation.[2]

***AI's entrance into Go.*** Demis Hassabis, head of the Google DeepMind team, noted that "Go is the most complex and beautiful game ever devised by humans … the richest in terms of intellectual depth" (Knight, 2016). Go has about $250^{150}$ possible moves, and the search space is often described as "a number greater than there are atoms in the universe" (Silver et al., 2016).[3] The seemingly unlimited number of possible moves in Go cannot be exactly identified by brute force calculation (as supercomputers have done with chess); Around two decades ago in the past two decades, several Go software programs—such as GnuGo 2.0, Pachi, and Crazy Stone—were released, but the performance of these programs was far inferior to that of professional Go players who use superlative "intuition" and evaluation skills in making certain moves (Knight, 2016).

Even if the latest supercomputers cannot calculate all possible moves in Go, recent advancement in deep-reinforcement-learning algorithms has improved AI remarkably. Instead of evaluating all possible solutions, AI uses these algorithms to reduce the potential moves to be considered and predicts sequential outcomes and winning probabilities.[4] AlphaGo, the initial APG with these algorithms, was invented by Google DeepMind. After several quality tests, in 2016 Google held a historic Go match between AlphaGo and the human Go master, Lee Sedol. Prior to this match, Lee and other Go experts expected that Lee would sweep all five games. Yet AlphaGo beat Lee 4–1, "a feat previously thought to be at least a decade away" (Silver et al., 2016). This event has been described as one of the milestones in the history of AI (Press, 2021).

[2] Other examples of major competitions include the Nongshim Cup—the competition between Team China, Japan, and South Korea—which awards $450,000 to the winning team. The Ing Cup (also known as Go Olympics) is held every four years and awards $400,000 to the winning player. In 2020 Shin Jin-seo, a twenty-one-year-old from South Korea, earned $920,754 in award money; Imaya Yuta, a thirty-year-old from Japan, earned $1,179,456.
[3] For comparison, chess has about $35^{80}$ possible moves. After the first two moves, chess has 400 possible next moves, while Go has 130,000 possible next moves (Muoio, 2016).

AlphaGo's success shocked not only Go players but also the public, who believed computers to be far inferior at intuitive judgments made amid enormous complexity. The match unexpectedly demonstrated that an APG could surpass the best human player and completely changed how players learned and practiced Go; since Leela was publicly released in 2017, professional players have learned from such APGs as Leela Zero, KataGo, and Handol (Somers, 2018).[5]

***How much better at Go is AI compared to humans?*** Go players are ranked and evaluated using the Elo rating system.[6] Figure B.1 in Appendix B shows how Elo scores have evolved among Go programs. Non-AI Go software—GnuGo, Pachi, and Crazy Stone—scored less than 2,000. The best human players scored around 3,800. In contrast, the scores of recent APGs, which are based on deep-reinforcement learning, far exceed 4,000. Given this gap in Elo ratings, even top professional Go players have no chance of winning against APGs. Put differently, the moves suggested by APGs yield the highest probability of winning, and even the best player can learn significantly from APGs.

[4] The APG context, therefore, is distinct from the general development of IT; APGs are about high-dimensional calculations and predictions that only become possible with AI and deep-reinforcement-learning algorithms.
[5] Before AI, professional Go players learned from books and past games. They also held group discussions (e.g., post-match game reviews), but it was generally impossible to quantitatively analyze moves or games.
[6] The Elo rating is calculated based on the relative capabilities of two players and their game outcome. The system has been widely used in other sports such as chess, football, basketball, and soccer.

***Learning from APGs.*** In the game of Go, AI technology is employed as a learning tool. APGs are not designed to provide real-time predictions to players on the spot during professional matches (which is strictly prohibited) but rather are used as a superior instructional tool to enhance players' decision-making capabilities. This distinction is important, as it highlights the potential of AI not only as a productivity-enhancing tool but also as a means of experiential learning.

The testimonials of professional Go players highlight the learning effect of APGs, which they largely attribute to its superior performance. Shin Jin-seo (who was ranked first in the world in 2020) provides further insights (Noh, 2019):

> "I have been using an APG since 2017. … I look at the APG's suggested move(s) and review other positions. … An APG is also used to predict the moves of opponents in the early stages. … Comprehending the move-level winning probability offered by APG is the new way I learn."

Figure A.1 in Appendix A provides a practical example of the information that professional Go players obtain from an APG. At any point in the game, the APG displays several optimal moves along with the winning probability associated with each suggested move; the different color schemes of the suggested moves make it easy to distinguish the very best move from others. Further, as a player chooses a move, the APG displays the optimal responses to that move, helping that player predict the opponent's responses in the following move. Repeating this learning process enables professional Go players to substantially improve both their understanding of strategic interactions in the game and their decision-making skills.

### 3.2. Research design

We compare changes in the quality of moves by professional players around the first public release of an APG. In 2016, when AlphaGo was the first APG to beat the best professional Go player, only a scientific article about its algorithm—not the program itself—was available to players. The first *public* APG that outperformed the best human player was Leela; its February 2017 update utilized the deep-reinforcement-learning algorithm used in AlphaGo. A few months later, a new version, Leela Zero, was developed based on the algorithm of AlphaGo Zero; this version had substantial impact on professional players. For example, the Korea Baduk (Go) Association and the South Korean National Go Team used Leela Zero for learning and training.

Importantly, the development of APGs did not arise from demands of Go players. Before AlphaGo, Go programs could only play at the level of human amateurs, and professional players did not believe that they could ever be beaten by computer programs. DeepMind decided to develop the AlphaGo program solely because of Go's profound complexity (Burton-Hill, 2016). Furthermore, the developer of Leela, Gian-Carlo Pascutto, made it clear that, although he had no interest in playing Go himself, he wanted to understand how deep learning worked. AI's entrance into Go, therefore, is not correlated with preexisting conditions in the Go industry.

To estimate the impact of APGs on the quality of moves by professional Go players, we first use a time-trend analysis. The event of interest occurred in February 2017 when a major update of Leela adopted the deep learning algorithm similar to AlphaGo. We conduct the analyses at the player-game level. Our sample consists of major professional Go games held from 2015 through 2019.

We then conduct a version of difference-in-differences estimation to understand the differential effects of APG. In an ideal world, we want to observe individual-level APG usage over time; unfortunately, such data is not available. Alternatively, we identify different age groups and compare the effects for younger players ("treated") as opposed to older players ("comparison"). Although we do not have a clean control group—some players in the older group may have also adopted APG—we expect younger players to have adopted APG to a greater extent. Comparing the size of the relative effect for the younger and older groups will produce a smaller estimate (i.e., biased toward zero) than would an ideal estimation that uses a clean control group with no APG usage (see, for example, Agrawal et al., 2016; Kang & Lee, 2022; Lipsitz & Starr, 2022). We run a set of robustness checks and adopt a similar approach for a country comparison.

We focus primarily on early moves—the first thirty moves of each game—because, like in many other games, a great opening is critical to winning at Go. Chang-ho Lee, a once-in-a-century player, pointed out the importance of the opening and likened it to a blueprint for architecture; the opening strategies are general roadmaps to the way players lead the game (Noh, 2016). We also analyze later stages and compare the results in Section 5.5.

### 3.3 Data

***Go games and professional players.*** We collect data on professional Go games held from 2015 through 2019 from the Go4Go database, which has been widely used in studies of Go (e.g., Chao et al., 2018). The data contains detailed information on each game, its players, Komi (the number of bonus points given to the player who moves second), the sequence of all moves, and the game outcome. From Go Ratings we gather additional data on the ages, nationalities (e.g., Chinese, Japanese, South Korean, Taiwanese, and others), and annual rankings of professional players. We multiply negative one by the ranking and divide it by 1,000 to ease the interpretation of the result; the higher the value, the better the player. To control for the difference in players' capabilities for each game, we create a variable, *Rank difference*, as the difference between the raw rankings of two players; we divide this difference by 1,000 such that a positive value indicates that the focal player's ranking is lower than the opponent's ranking.

***Measuring the quality of moves.*** Since Leela Zero provides the probability of winning for any possible move made at any particular point of the game, we use it to calculate the difference in winning probability between a move made by a professional player and Leela Zero's suggested move, a move that would achieve the highest winning probability among alternative moves. Our main dependent variable is $Move\ Quality_{ig}$, which represents the average difference in the winning probability of focal player *i*'s move compared to the APG's corresponding solution for the first thirty moves of a game *g* (i.e., the game's 1st, 3rd, 5th, …, 29th moves if the focal player moves first or the 2nd, 4th, 6th, …, 30th moves otherwise). For each game, we calculate separately the value of the move quality for each player *i* (i.e., the black stone holder and the white stone holder):

$$Move\ Quality_{ig} = \frac{\sum_{n=1}^{15}\left(\begin{matrix}The\ winning\ prob.of\ the\ focal \\ player\ i's\ n^{th}\ move\ in\ a\ game\ g\end{matrix} - \begin{matrix}The\ winning\ prob.of\ the\ APG's \\ solution\ to\ the\ move\ in\ a\ game\ g\end{matrix}\right)}{15}$$

where *n* represents the order of the focal player's move. $Move\ Quality_{ig}$ takes a *non-positive* value (since APG is superior) and ranges from –100 (lowest quality) to 0 (highest quality). A smaller absolute number indicates a higher-quality move by the player. If a player places stones as suggested by the APG for all moves, the average difference in winning probability between the player and the APG is zero (*Move Quality*=0). This variable becomes larger in absolute value as a player's moves worsen, that is, as they deviate from the best moves suggested by the APG.

We used Leela Zero (May 23, 2020 version) along with the GoReviewPartner program to analyze all 749,190 moves in 24,973 games played from 2015 through 2019. The computation took about three months; Appendix A.2 provides the calculation and implementation details.

***Summary statistics.*** Table 1 provides descriptive statistics on the key variables at the player-game and player levels. Table 1(a) includes two observations for each game: one for the first mover (black stone holder) and another for the second mover (white stone holder). After omitting games that lacked information on players' ages or ranks, our final sample has 46,454 observations. The mean of our main dependent variable, $Move\ Quality_{ig}$, is –2.01 over the sample period. That is, the players' winning probability for the first thirty moves in a game averaged 2.01 percentage points less than that of the APG's best move. This is a substantial difference because the difference of two percentage points for each move accumulates as the game progresses. The average (raw) rank of the players is 280th before transformation. The average rank difference is, by definition, zero (the positive and negative differences of the two players cancel each other).

Table 1(b) shows the descriptive statistics at the player level. We identify 1,241 players from 2015 through 2019. The average age of players is 32.41, and the median age is 26.98.

*"Insert Table 1 here"*

## 4. RESULTS

### 4.1. Did APG improve the quality of moves by professional players?

***Model-free evidence.*** We first graphically present our main outcome of interest. Figure 1(a) shows the weekly average value of *Move Quality*$_{ig}$ from 2015 through 2019. The vertical line indicates February 2017, the date of the public release of Leela, the first APG that surpassed human performance. This model-free illustration shows that while *Move Quality*$_{ig}$ was relatively low and stable over time before APGs, it increased immediately after Leela's public release.

*"Insert Figure 1 here"*

***Time-trend analysis.*** We then use a formal OLS regression model to estimate the *Move Quality*$_{ig}$ of professional Go players around the first public release of an APG in February 2017. The baseline time-trend regression specification at the player-game level is $Y_{ig} = \alpha + \beta_1 \cdot Post_g + \gamma_i + \delta_{-i} + \epsilon_{ig}$, where indices $i$ and $g$ represent player and game, respectively. Focal-player-fixed effects are represented by $\gamma_i$, while $\delta_{-i}$ represents fixed effects for the opposing player. $Y_{ig}$ is *Move Quality*$_{ig}$. $Post_g$ is equal to 1 if a Go game is played in a quarter after February 2017 (when the APG was released) and 0 otherwise. Standard errors are clustered at the focal-player level to address a concern that the error terms are correlated across the players. $\beta_1$ captures how APGs improved the move quality of players.

The results are shown in Table 2. Column 1 shows that the coefficient of $Post_g$ is positive ($\beta$=0.756, $p$<0.01), indicating that *Move Quality*$_{ig}$ increased by 0.756 percentage points (or about

30.5 percent) on average after the APG's public release.[7]

*"Insert Table 2 here"*

It is possible that the performance of professional players had been improving over time and that such improvement drove the results, although Figure 2 does not indicate evidence of this. To control for this trend, we add a $Trend_g$ variable (i.e., the number of quarters elapsed since the first quarter in our sample) and an interaction term ($Post_g \times Trend_g$). The results are shown in column 2. We find a small yet positive trend ($\beta$=0.007, $p$<0.05), suggesting that the performance of professional players had improved slowly over time. Importantly, the coefficient of the interaction term ($\beta$=0.116, $p$<0.01) shows that there are much larger—that is, about seventeen times greater—improvements following the public release of the APG, even after performance trends are taken into account. The effect in the 10th quarter (i.e., the first quarter after the APG release) is 0.222 (–1.007+0.007×10+0.116×10).

**4.2 Did the improvement stem from learning from AI?**

To support our argument that such an improvement stems primarily from learning from AI, we further analyze the *Move Match*, or the degree to which each move aligns with the APG's best solutions. If players have learned from APGs, the likelihood of their making exactly the same moves as the APG's top suggestions should increase. Given that APGs are not available during gameplay, a player's moves that match those of the APG exactly indicate that the player learned from the APG and improved their skills prior to the game. An indicator variable, $Move\ Match_{ig}^{k}$, captures, on average, how many moves of the focal player $i$ are the same as the APG's top $k$ suggestions among the first thirty moves in a game $g$. We consider $k$=1 to be an exact match between the player's move and the APG's top suggestion. If $k$=3, we check whether the player's

[7] All percentage changes are calculated as the relative changes in the average move quality of games played by the players of interest during the quarters of the sample period before the release of the APG.

move is among the APG's top three suggestions:

$$Move\ Match_{ig}^{k} = \frac{\sum_{n=1}^{15} \mathbf{1}\left(Player\ move_{ing} \in \left\{APG\ move_{ng}^{1}, APG\ move_{ng}^{2}, \ldots, APG\ move_{ng}^{k}\right\}\right)}{15}$$

Figure 1(b) illustrates the weekly average value of $Move\ Match_{ig}^{k}$ for APG's top suggestion ($k$=1). Before APGs, $Move\ Match_{ig}^{k}$ remained low and unchanged over time, but after Leela's public release it notably increased. This result is consistent for APG's top three ($k$=3) and top five ($k$=5) suggestions and consistent when we conduct time-trend analyses (see Table B.1 of Appendix B), suggesting that players' improvements are indeed the result of learning from AI.

### 4.3. Differential effects of AI adoption and utilization by age

As discussed in Section 2.2, age is an important factor that can affect the adoption and utilization of new technology. We plot in Figure 2, Panel (a) the model-free illustration of two different age groups: Young and Old. This figure shows that $Move\ Quality_{ig}$ was relatively stable and similar among the two groups before the APG release, while the increase in $Move\ Quality_{ig}$ is notably greater for the Young group afterward.

*"Insert Figure 2 here"*

We then formally test whether the APG indeed has differential effects on the move quality of professional players of different ages. We estimate the following model at the player-game level:

$$Y_{ig} = \alpha + \beta_1 \cdot Post_g \cdot Young_i + X_{ig} + \gamma_i + \delta_{-i} + \theta_g + \epsilon_{ig},$$

where $\gamma_i$, $\delta_{-i}$, and $\theta_g$ represent focal-player-, opponent-player-, and quarter-fixed effects, respectively, for game $g$. $X_{ig}$ includes control variables such as *Komi*, *White*, *Rank*, and *Rank differences between players* at the player or game levels. $Young_i$ is an indicator variable equal to 1 if the player's age is less than the median age of all players (i.e., less than twenty-seven years) as of Leela's public release in February 2017, and 0 otherwise.

Table 3 shows the results. Column 1 includes only $Young_i$ and control variables with quarter-time-fixed effects. Column 2 then adds the interaction term, $Post_g \times Young_i$. The coefficient of the interaction term ($\beta$=0.268, $p$<0.01) is positive; the quality improvement among younger players is 0.268 percentage points (or 10.9 percent) greater than that for older players.

To check whether our results are robust when players' inborn characteristics are considered, we add the player-fixed effect in column 3 and the opponent-player-fixed effect in column 4. The effect of AI is consistently more prominent for the younger group, whose quality of moves improved by 0.203–0.268 percentage points (or 8.2%–10.9%) over that of the older group.[8]

*"Insert Table 3 here"*

We again consider *Move Match* to substantiate our argument that differential improvements by age are indeed driven by players' learning from APGs. Table 3, columns 5–7, shows the results from estimations with $Move\ Match_{ig}^{k}$ as the dependent variable. After the APG, younger players ($Post_g \times Young_i$) were more likely than older players to make moves that matched the APG's top suggestions. What is more interesting and convincing is that the estimates shrink as we broaden the set: 0.031 ($k$=1; column 5), 0.025 ($k$=3; column 6), and 0.018 ($k$=5; column 7). This occurs because, when players learn from an APG, they are more inclined to learn the very best move ($k$=1) than the near-best ones; as we expand the set ($k$), the impact of AI thus diminishes.

**4.4. Robustness checks**

We further check the robustness of the results in six ways: 1) an estimation with distributed leads and lags, 2) a sensitivity analysis by age conditions, 3) an analysis using monthly data, 4) the different numbers of moves for opening strategies (the first 15, 20, 40, 50, or 60 moves), 5) an analysis of earlier Go programs, and 6) players' nationalities as a proxy for APG exposure.

[8] Note that the magnitude of the effect is smaller than that in the main analysis (0.756 percentage points in Table 2, column 2). As discussed in Section 3.2, this is because our empirical design uses older players as the comparison group; this group was also affected by APGs in the same way as the younger group (although to a lesser extent).

***Estimation with distributed leads and lags.*** To check the pre-APG trend and the time-varying effects of the APG, we include the distributed time leads and lags in our regression:

$$Y_{ig} = \alpha + \Sigma_z \beta_z \times Z \times Young_i + X_{ig} + \gamma_i + \delta_{-i} + \theta_g + \epsilon_{ig},$$

where $\gamma_i$, $\delta_{-i}$, and $\theta_g$ represent focal-player-, opponent-player-, and time-quarter-fixed effects, respectively. The symbol $Z$ represents the indicators for time leads and lags—that is, the number of quarters before or after the public release of the APG.

Table B.2 of Appendix B, columns 1 and 2, shows the detailed regression results, and Figure 2(b) graphically illustrates the results. We do not find any pre-APG trend for $Move\ Quality_{ig}$; the estimates for pre-APG quarters are close to and statistically not distinguishable from zero. For quarters after the APG's release, the improved quality among younger players is large and persistent.

***Sensitivity analysis for age groups.*** We examine whether the results are sensitive to our operationalization of age groups. First, we use the average age (instead of the median age) as the cutoff for the younger and older groups; this increases the cutoff age from twenty-eight years to thirty-three years. The results in Table B.3 of Appendix B are robust to this alternative classification ($\beta$=0.268, $p$<0.01 in column 4). Second, we investigate the same model with three age groups based on the age tertile: Young (bottom tertile); Middle (middle tertile); and Old (top tertile). The results are provided in Table B.4 of Appendix B. The estimates for $Post_g \times Young_i$ ($\beta$=0.338, $p$<0.01 in column 4) and $Post_g \times Middle_i$ ($\beta$=0.248, $p$<0.01 in column 4) are large. Importantly, the effect is most pronounced when Young players are compared to Old players, and the magnitude is smaller for Middle players. We obtain similar results when classifying players' ages into three categories: under age twenty; twenties (ages 20–29); and thirty or older.

***Alternative time-fixed effects.*** To consider the time effect on a more granular level, we estimate the model with month-fixed effects instead of quarter-fixed effects. Table B.5 of Appendix B shows that the results are consistent with this alternative. Figure B.2 of Appendix B graphically illustrates the results obtained from the models with the distributed time leads and lags at the month level; these results are similar to those shown in Figure 2(b). We once again confirm the parallel time trend before the release of the APG and the substantial effect post-APG.

***Opening strategy with different numbers of moves.*** Our results could have been influenced by the choice of the number of moves. To check this possibility, we estimate our models with different definitions for early opening moves: the first 15, 20, 40, 50, and 60 moves of the game. The results, shown in Table B.6 of Appendix B, are robust to these alternative definitions.

***The effect of earlier Go programs.*** Although Go programs prior to AlphaGo or Leela did not perform at the level of top human players, these programs may have offered training opportunities for professional players similar to the ways that training sessions with early chess computers have been shown to improve the skills of chess players (GP, 2023). The introduction of earlier Go programs therefore provides a valuable opportunity to check whether the effects are driven by learning from APG's superior performance or by more frequent training opportunities (albeit with an inferior performance). We examined the impact of the earlier Go program, Crazy Stone, released in 2015; its Elo rating, just below 2,000 was inferior to the best human level (around 3,800; see Figure B.1 in Appendix B). Figure B.3 in Appendix B illustrates the results. We do not find any improvement in move quality after the release of Crazy Stone. This result rules out the possibility that more frequent training sessions (with inferior programs) are driving the findings and supports the proposition that learning from AI (i.e., from superior APGs) is the key channel through which players have improved their move quality.

***Players' nationalities as a proxy for APG exposure.*** A Go player's exposure to AI may differ depending on the player's nationality. Of the three major countries with the largest professional Go leagues, access to or exposure to APGs has been relatively lower in Japan. For example, no AlphaGo match was held in Japan. Further, the interest score in *AlphaGo* from Google Trends was 4 in Japan compared to 100 in China and 92 in South Korea. In an interview, an expert explained that, in Japan, Go is considered an art form, which partly explains its slower adoption. We estimate a version of a difference-in-difference model comparing players in countries significantly affected by APGs (i.e., China and Korea) with those in a country less affected (i.e., Japan). We find that move quality improved significantly more for the former (see the Appendix C for details).

## 5. FURTHER ANALYSES

### 5.1. How did players improve when their opponents used AI?

So far, we have focused on the focal player. An important question is, how does the quality of player moves vary by the extent to which opposing players learn from AI? If learning from AI indeed drives move-quality improvement, the effect should be greater for player pairs when both have heavily utilized AI; this is because play between such players most resembles situations where both have learned from AI. To check this, we split the sample by player age and by country and conduct a series of time-trend analyses to examine variation in the effect across different dyadic pairs. The results are illustrated as a heatmap in Figure 3. The move quality has improved across all pairs, but the effect is particularly marked among pairs of Young versus Young. In contrast, the improvement is relatively smaller for pairs of Old versus Old. This finding—that the effect is magnified when the moves of a player's counterpart are more likely to be similar to those of APG—once again bolsters our argument that players indeed learn from AI (see Figure C.3 of the Appendix C for a comparison by nationality).

*"Insert Figure 3 here"*

### 5.2. Did better players improve more?

The increase in average-move quality does not necessarily mean that players of different skill levels improved to the same extent. To explore this, we first examine the effects across the distribution of players' skill levels (à la Athey & Imbens, 2006; Lipsitz & Starr, 2022). By employing the change-in-changes method, we estimate the quantile treatment effects of APG by comparing the move quality of younger players with that of older players at different points in the player-game level skill distribution. The effects, illustrated in Figure 4, are positive across the entire range of the distribution, and the effect size is greater for those at the bottom of the skill distribution.

*"Insert Figure 4 here"*

Further, we compare the improvement over time of players in the top decile (10th decile) with those in the bottom (1st and 2nd) deciles. The model-free evidence is illustrated in Figure B.4 of Appendix B. Panel (a) shows improvement in move quality, even among the top decile. Further, in Panel (b), while the top performers demonstrated notable improvement (up to 44.8%), the increase was even more prominent for players in the bottom deciles (up to 49.3%). This indicates that the performance gap between top players and others narrowed after introduction of the APG.

### 5.3. Mechanisms for quality improvement: Errors and critical mistake

We extend the analysis beyond *Move Quality* and delve into two important channels through which AI-based learning improves the quality of moves: *errors* and *critical mistakes*. This analysis is motivated by the norm that, after completing a Go game, players spend significant time and effort analyzing and evaluating each move—especially if a move was an error or a critical mistake. In an interview, Shin Jin-seo (who was ranked first in the world in 2020) stated:

> *Before APG, players and their peers replayed games and discussed which move was an error and which was a critical mistake. After the public release of APGs, this replay and discussion by players became almost meaningless. APGs teach us by showing the accurate winning probability with each move. If the winning probability drops from 60 percent to 40 percent after a move, that move is an error. If the probability drops from 80 percent to 20 percent, that is a critical mistake. ... I have to admit that APG-based training provides limitless help in developing my Go skills* (Sohn 2021).

To test these mechanisms, we measure the error in a game as the number of "bad" moves, those in which the winning probability drops by 10 or more percentage points when compared to the winning probability of the focal player's immediately preceding move. The "critical mistake" is the magnitude of the biggest drop in winning probability among all the moves in a game. Figure B.5 of Appendix B shows the model-free trend of errors (in Panel a) and the critical mistake (in Panel b). Both the errors and the critical mistake decrease substantially after the release of the APG.

We then conduct regression analyses on errors and the critical mistake. Table 4, columns 1 and 3, shows that the number of errors and the magnitude of the critical mistake decreased after APG release. Columns 2 and 4 show the results after controlling for the linear trend. The estimates for the interaction term ($Post_g \times Trend_g$) show that the (preexisting) negative trend ($\beta$=–0.009, $p$<0.01) is discontinuously accelerated after the introduction of APG ($\beta$=–0.233, $p$<0.01). These results confirm that learning from AI improved the quality of moves of professional players by reducing both the number of errors (33.7%) and the magnitude of the critical mistake (21.9%).

*"Insert Table 4 here"*

### 5.4. Did AI-driven improvements in move quality lead to winning?

Building upon our finding that younger players improve more than older players after learning from APG, we further investigate whether this improvement leads to a higher probability of winning a game. A model-free comparison of means reveals that the winning rates of younger players improve when they play against older players. The average winning rate of younger players increased by 10.2%, from 53.1% pre-APG (2015–2016) to 58.5% post-APG (2017–2019). We then formally conduct the three-step mediation analysis suggested by Baron and Kenny (1986). As a baseline model, we run the linear probability model of winning a game on an interaction between an indicator for a younger player and an indicator for a post-APG period. In Table 5, column 1, the improvements in move quality indeed have led to a higher chance of winning ($Post_g \times Young_i$: $\beta$=0.024, $p$<0.01); the chances of younger players' winning average 2.4 percentage points (4.6%) higher post-APG, if other variables are set to mean values.

We then conduct the mediation analysis to test for the channels. First, we check whether $Post_g \times Young_i$ is statistically related to the proposed mediators: *Move Quality*, *Errors*, and *Critical Mistake*. Table 5, columns 2–4, shows that *Move Quality* is positively associated with the younger group after APG, while *Errors* and *Critical Mistake* are negatively associated. Second, we confirm in columns 5–7 that *Move Quality* is positively associated with the probability of winning but negatively associated with *Errors* and *Critical Mistake*, without the explanatory variable ($Post_g \times Young_i$). Third, we examine whether the magnitude of the estimated effect of the explanatory variable ($Post_g \times Young_i$) decreases with inclusion of the mediators. In columns 8–11, the estimates for the explanatory variable ($Post_g \times Young_i$) are shown to decrease for all cases after adding the mediator variables when compared with those in the baseline model (column 1). In the separately estimated mediation models for the three mediators—*Move quality*, *Errors*, and *Critical Mistake*—the indirect effects through the explanatory variable account for 18.9 percent ($p<0.01$ in the Sobel test), 4.2 percent ($p \approx 0.01$), and 13.1 percent ($p<0.01$) of the total separate mediation effects, respectively. Taken together, younger players were more likely to win after the APG release through their improvements in three dimensions: *Move quality*, *Errors*, and *Critical Mistake*.

*"Insert Table 5 here"*

**5.5. How did the AI effect vary throughout the game?**

Although we focus on the early (first to thirtieth) moves in the main analyses, AI's role is not restricted to this particular phase. Here we extend the analysis to include later stages of the game, incrementally adding thirty moves (up to 180 moves) to our analysis. We graphically present model-free results on $Move\ Quality_{ig}$ in Figure B.6 of Appendix B. The AI effect is most prominent in early opening moves (for moves 1–30) and gradually decreases as we include later moves in the analysis. Formal analyses confirm these observations. Table B.7 in Appendix B shows the results from six different regression specifications. The estimate for $Post_g \times Young_i$ gradually shrinks from 0.203 (for moves 1–30) to 0.050 (for moves 1–180). The estimates with distributed leads and lags are graphically illustrated in Figure B.7 of Appendix B; the improvement among younger players is highest for the opening strategy and weakens as moves from later stages of the game are included.

One explanation for this may be uncertainty. At the early stage of a game, when only a few stones have been placed, players have the highest number of possible moves, and their ability to assess all alternatives and subsequent moves is significantly limited. In other words, prior to APG training, players relied more on heuristics or conventional opening strategies to alleviate such uncertain environments in which complete evaluations are not possible. This is where learning from AI can most help players to improve the quality of moves. As a game progresses into its mid-to-late stages, uncertainty is reduced as more stones are put on the board, and it becomes less difficult to evaluate potential moves. The results suggest that the learning effect from AI can vary depending on the uncertainty of the environment and the opportunity to learn from AI.

## 6. DISCUSSION AND CONCLUSION

### 6.1. APG and chess computers

The focus of this study aligns with GP's central question regarding the impact of chess computers on players' performance. GP leveraged a differential access to chess computers based on players' location (Western Europe versus Soviet Union) and analyzed 20,000 chess players in 500,000 tournaments that took place during 1970–2000. The emergence of chess computers offered an initial testbed for studying the impact of AI-assisted training in complex strategic interactions. GP find a positive effect of the introduction of chess computers on players' performance. Importantly, the benefits of chess computers were confined to players whose skills were inferior to the chess computers, suggesting a catch-up effect along the skill distribution. GP then argues that training opportunities are the primary mechanism; that is, chess computers are a substitute for human training partners, which are limited in supply, although the substitution is not complete because, unlike humans, computers do not make mistakes.

Extending GP's research, our study brings unique insights into how humans learn from AI and improve their decision-making. First, the APG is powered by advanced (deep) reinforcement learning. Unlike the chess computers of the 1970s–1990s, APG performance has surpassed that of the best human players.[9] Second, APGs were released free of charge and nearly simultaneously, with minimal gaps in their release times, which provides a unique opportunity to assess the impact of AI penetration.[10] Third, APGs provide players with more information. An APG makes three to five suggestions, gives the winning probabilities associated with those moves, and further suggests the likely next five to ten moves. Chess computers, in contrast, provided only one deterministic move and no further information.

Fourth, taking advantage of APGs' superiority, we study how humans *learn* from AI; this takes a step forward from GP's view of AI as a training (sparring) partner that is not necessarily better than the trainee. Our study also highlights the democratization of high-quality learning opportunities, as players are able to learn from the strategies and decisions of the very best player, APG. The APG's superior performance was paramount in enhancing the skills of professional Go players, perhaps because Go is much more complex than chess. Fifth, we expand our discussion on the boundary conditions—namely, the ages and skill levels of players, exposure to AI by country, and the stage of the game—of AI's instructional roles. We also conduct dyadic analyses by players' age and country to study interactions between players. We hope this scholarly dialogue with GP deepens the understanding of how humans train with and learn from AI.

[9] The Elo rating of a pre-APG software, Crazy Stone, is similar to that of a later-stage chess computer in GP's study. Unlike the findings of GP, we do not find evidence that Crazy Stone improved the move quality of Go players (see Section 4.4).

[10] In contrast, chess computers came at a relatively high cost and were gradually diffused over several decades. For instance, the first commercial chess computer in 1977 cost $200 (equivalent to approximately $1,000 today).

### 6.2. Implications and limitations

Our findings may not represent all types of human-AI interactions and their consequences. For example, AI might replace humans in certain tasks or domains rather than enhance human skills. However, instances where humans learn from AI and continue performing tasks themselves are neither rare nor unusual. The insights into learning from AI provided by professional Go games offer timely implications for the expanding role of AI and its relationship with humans.

First, AI could reveal that what humans have long considered to be solutions may not be the best approaches. AI has the potential to bring breakthroughs in human knowledge, heuristics, and routines and to pave the way for new paradigms (Choi et al., 2023). Second, AI has broader applications than merely substituting for or assisting with human tasks. We provide new theoretical and empirical accounts of how AI instructs and transforms human decision-making (Brynjolfsson et al., 2021). Although researchers have recently expanded their interest in AI's role in supporting human judgment (Choudhury et al., 2020; Kleinberg et al., 2017; Wang et al., 2019), studies have focused on AI's real-time role as an assistant, boosting task-related performance. We highlight the instructional role of AI, emphasizing its potential to improve human skills and performance. Our findings can be applied to domains where AI has already outperformed or will soon outperform human activities. For example, AI's performance in radiology rivals that of trained radiologists in triaging chest and breast x-rays and detecting lung cancers. Doctors learn from AI's analysis to provide better diagnoses and predictions (Grady, 2019; Lebovitz et al., 2021, 2022; Reardon, 2019). Third, not everyone may benefit from AI equally. We show that openness to new technologies and the ability (characterized by individual's age, experience, or cultural background) to utilize these technologies can influence the benefits reaped from AI. These findings contribute to the growing literature on the differential effects and potential inequality implications of AI (e.g., Beane & Anthony, 2023; Choudhury et al., 2020; Miric et al., 2020). Fourth, the impact of AI also depends on the complexity and uncertainty of a situation. In Go, AI-driven improvement is most prominent in the early stages of a game. This boundary condition of AI's effect is consistent with the findings in drug discovery and development (Lou & Wu, 2021) which suggest that a uniform application of AI would not yield optimal outcomes and could lead to inefficient allocation of AI and human resources. Therefore, careful consideration is required to determine where and to what extent AI should be adopted and utilized.

Several of our findings and implications are relevant to managers and organizations. In the era of rapidly evolving AI-related technologies, a central question facing firms is how to utilize AI to achieve a competitive advantage and enhance performance. Our study emphasizes the instructional role of AI and provides insights into its potential impact in the workplace. Workers can benefit from AI-powered learning programs, which can enhance their decision-making abilities, leadership skills, and strategic thinking in complex and uncertain business environments. For instance, this study highlights the promise of AI algorithms in workforce education and AI-driven human resource management, addressing issues such as algorithmic aversion and vintage-specific skills (e.g., Choudhury et al. 2020; Gaessler & Piezunka 2023; Krakowski et al., 2022; Tong et al., 2021). In the realm of talent acquisition and management, obtaining information about candidates beyond their cover letters and resumes is crucial for assessing their suitability. AI can provide accurate, personalized predictions on churn, performance, and suitability, and can identify key evaluation criteria. However, HR managers still conduct interviews, combining their knowledge and experience with insights learned from AI's analysis. This approach helps avoid losing potential talent due to algorithm aversion and better demonstrates the firm's commitment to them through human emotional intelligence. Other examples of successful application of AI include performance feedback, investor relations management, and fighter pilots in air force units (see the Appendix D for details).

In integrating AI tools and offering valuable learning opportunities to workers, it is critical for firms to be aware of boundary conditions and heterogeneity in tailoring these programs. For example, the gains from using AI tools are not uniform across all workers, and relatively low-skilled workers could gain more. When facing budget constraints, managers may prioritize using AI tools to enhance the performance of lower performers. Understanding how AI's effectiveness is influenced by an individual's age and prior exposure to AI could also lead to successful adoption and utilization of AI, fostering a firm's innovation and growth.

There are several limitations to this study. First, we do not have a clean control group. There does not exist a group without exposure to APG. Although we use player characteristics to proxy for APG adoption and utilization to mitigate these issues, both the treated and comparison groups are exposed, and thus our estimates provide the lower bounds of the effect. Relatedly, it is not possible to obtain individual-level data on APG usage. To mitigate this concern, we conducted auxiliary empirical analyses on move match and found that the improvement in move quality coincided with the increase in move match between players and APG. Second, the application of our findings to different contexts requires careful consideration. Although the domains where AI outperforms humans have broadened to include various settings—such as hospitals (Cadario et al., 2021), law firms (Kahn, 2020), and sports teams (Zarley, 2021)—the Go context presents unique characteristics and strategic dynamics that may differ from other domains. Despite these limitations, we hope this study enriches discussions on various aspects of AI, particularly learning from AI, and stimulates further research.

**Figure 1**. Effects of APG on average *Move Quality* and *Move Match*: Model-free evidence

(a). Move Quality

(b). Move Match (*k*=1)

*Note*. This figure illustrates the weekly average *Move Quality* (Panel a) and *Move Match* (Panel b) of players from 2015 through 2019. The gray solid line represents the raw (unprocessed) weekly average value. The blue solid line and the blue area around it show the smoothed trend (loess; span=0.7) and the 95% confidence interval, respectively. The vertical line represents February 2017, the date of the first public release of an APG, Leela.

**Figure 2**. Differential effects of APG on move quality by player age

(a) Model-free evidence

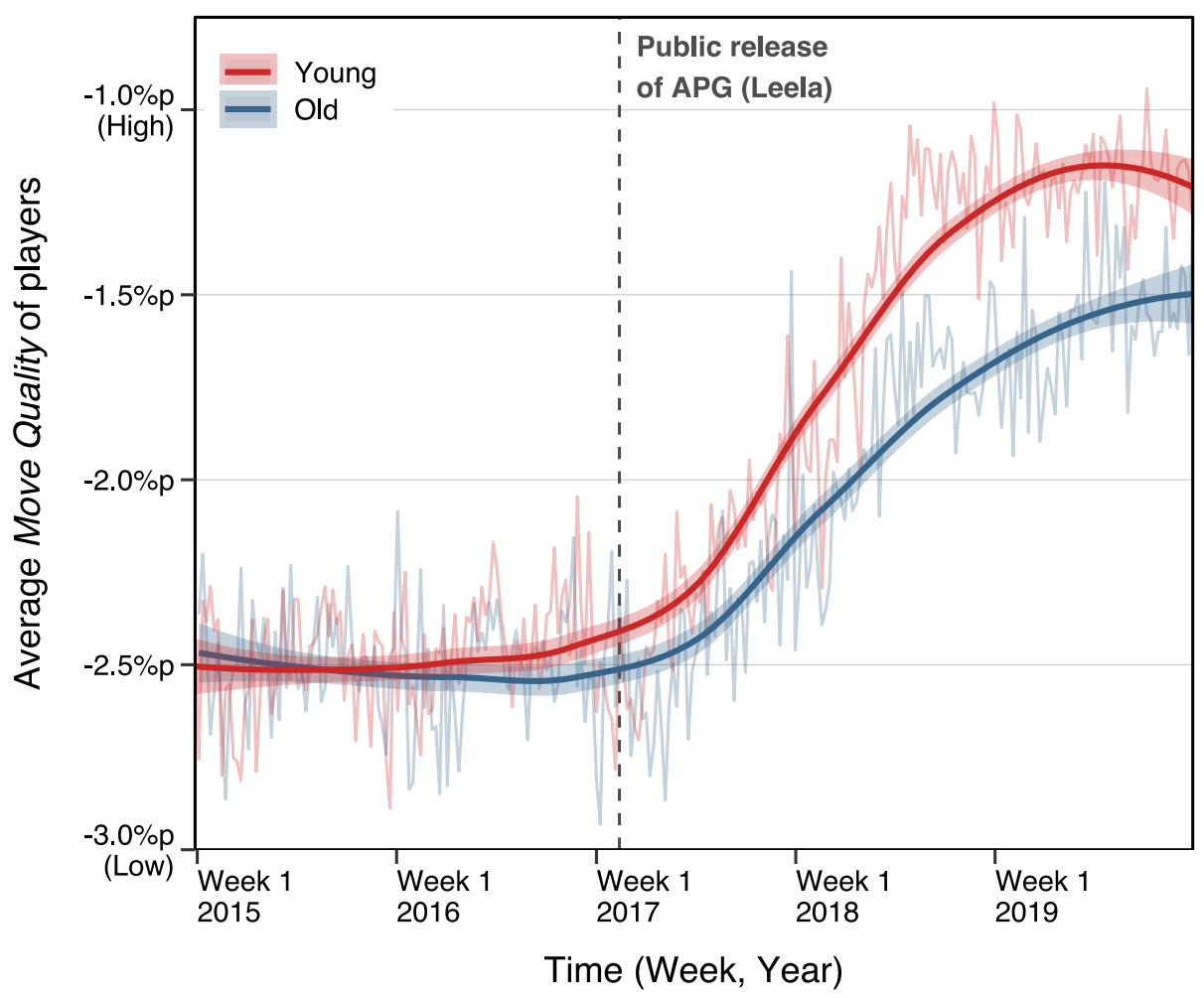

*Note*. This figure illustrates the average *Move Quality* of professional players by player age. The red and blue fluctuating lines show the raw (unprocessed) weekly average values for younger players (below median age) and older players (above median age), respectively. The red and blue smooth lines and the shaded areas around them show the locally smoothed trends (loess; span=0.7) and the 95% confidence intervals. The vertical line indicates February 2017, the date of the first public release of an APG, Leela.

(b) Difference-in-differences approach

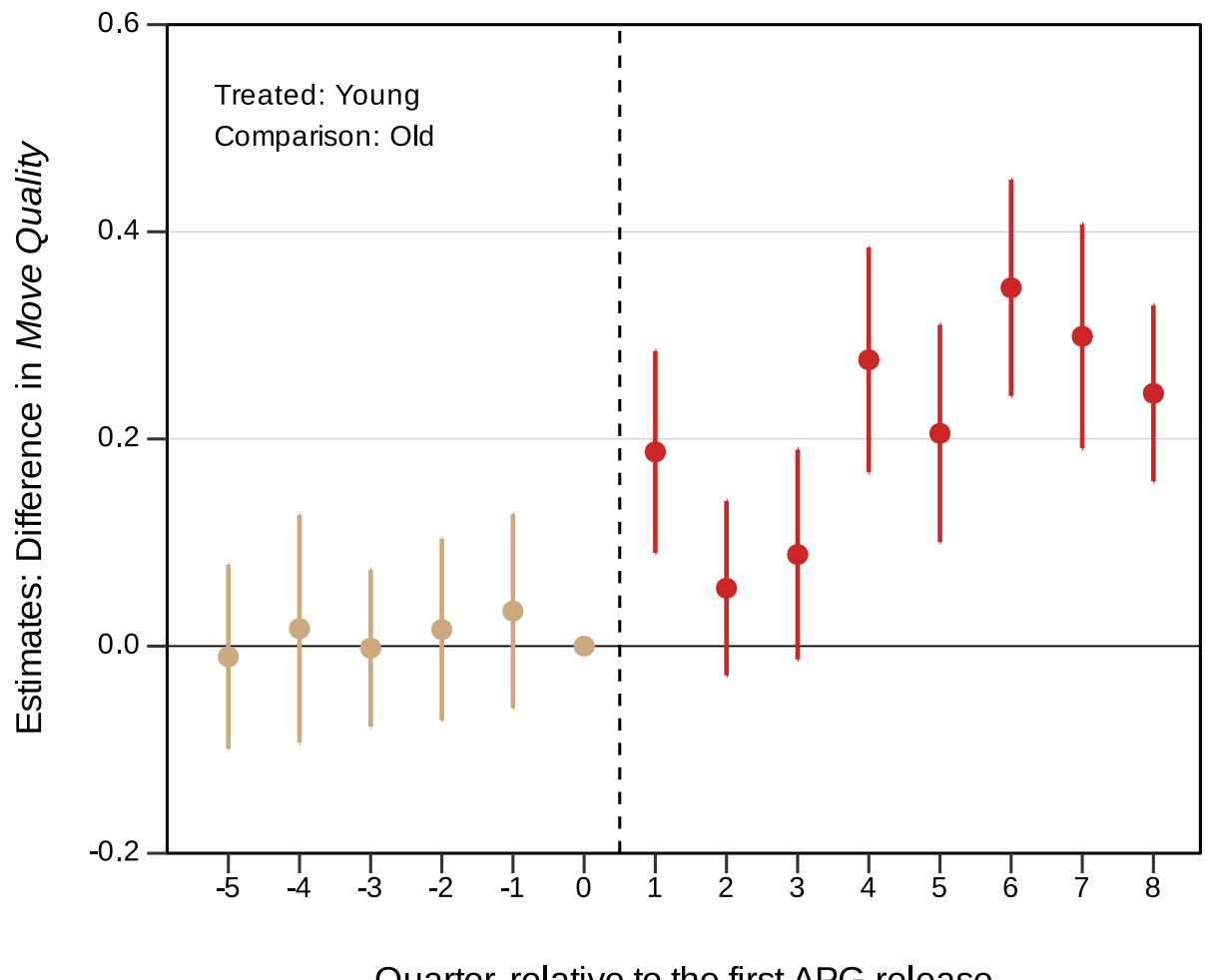

*Note*. This figure illustrates the differential effects of APG on *Move Quality* by player age. The points graphically present the *Move Quality* of younger players (those below the median age) compared to that of older players (above median age), based on the regression estimates in Table B.2 of Appendix B, column 2. The vertical error bars show the 95% confidence intervals. We do not find a difference in *Move Quality* by age before the APG release. After the APG release, the increase in *Move Quality* is greater for younger players than for older players.

**Figure 3**. The effects of APG on *Move Quality* across dyadic relationships of age group

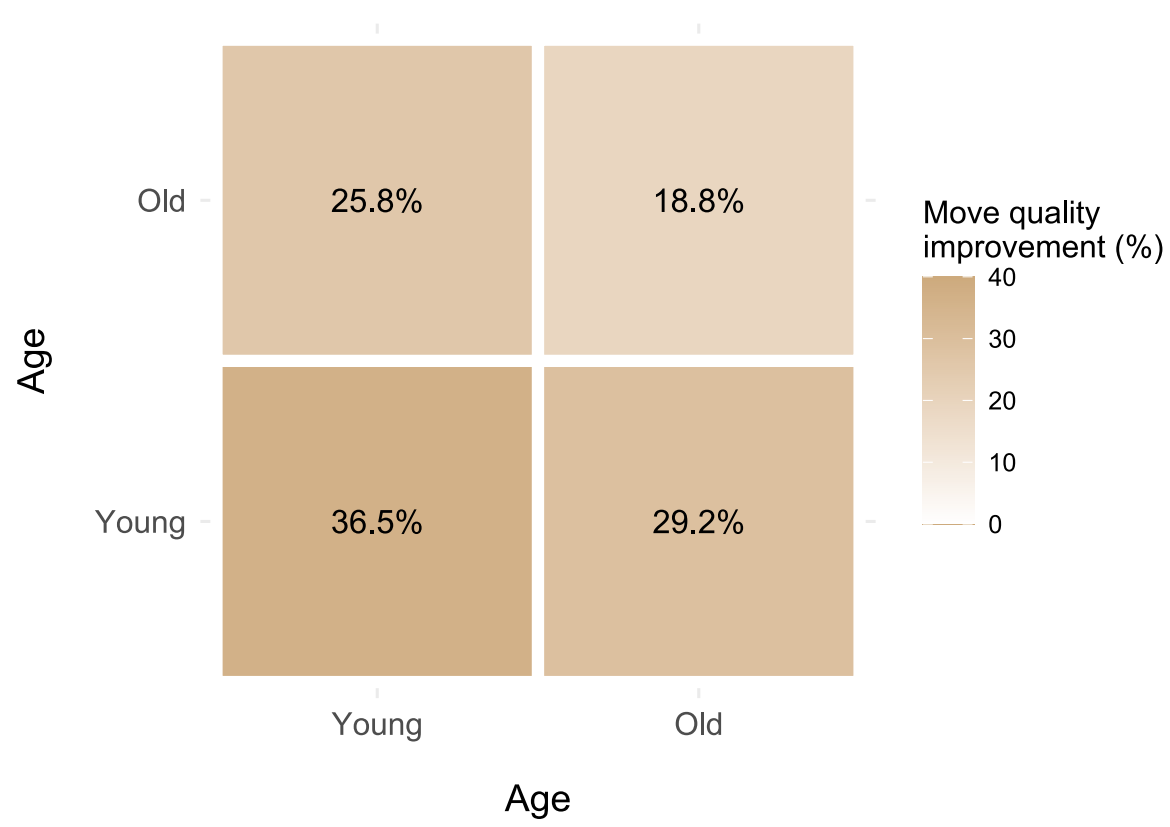

*Note*. This figure illustrates the improvement in *Move Quality* across various dyadic relationships, categorized by pairs of age groups, around the introduction of APG. The x-axis represents the focal player's age group, while the y-axis represents the counterpart's age group.

**Figure 4**. Quantile treatment effects on *Move Quality*

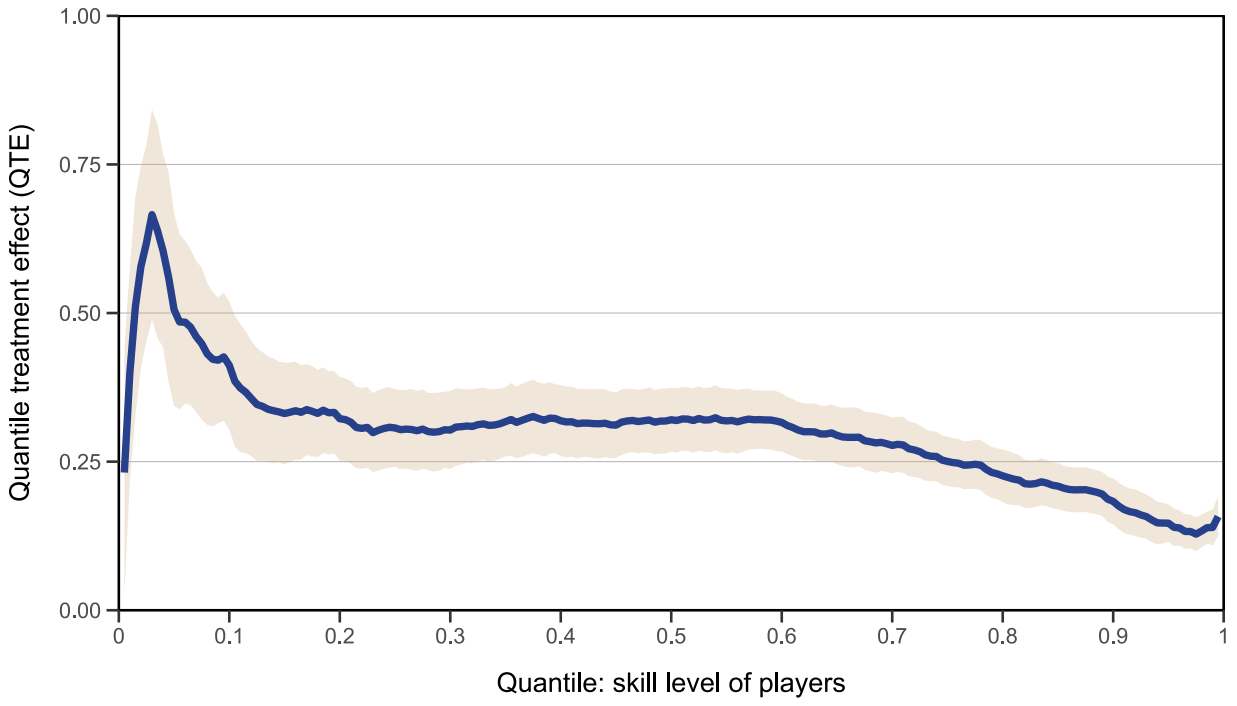

*Note*. This figure illustrates treatment effects at quantiles of *Move Quality* before and after the first public release of an APG, Leela, following the method suggested by Athey and Imbens (2006). The shaded areas show bootstrapped 95% confidence intervals.

**Table 1.** Descriptive statistics

(a). Player-game level

| | N | Mean | Median | SD | P25 | P75 |
|---|---|---|---|---|---|---|
| Move Quality | 49,946 | –2.01 | –1.92 | 1.07 | –2.66 | –1.22 |
| Number of Errors | 49,946 | 0.13 | 0.00 | 0.37 | 0.00 | 0.00 |
| Magnitude of the Critical Mistake | 49,945 | 5.66 | 4.80 | 3.96 | 2.94 | 7.36 |
| Age | 49,613 | 28.07 | 24.31 | 12.58 | 19.55 | 31.52 |
| Young | 49,613 | 0.62 | 1.00 | 0.48 | 0.00 | 1.00 |
| Rank | 48,813 | –0.27 | –0.17 | 0.27 | –0.42 | –0.05 |
| Rank Diff | 47,826 | 0.00 | 0.00 | 0.21 | –0.09 | 0.09 |
| White | 49,946 | 0.50 | 0.50 | 0.50 | 0.00 | 1.00 |
| 7.5 Komi | 49,946 | 0.38 | 0.00 | 0.49 | 0.00 | 1.00 |

(b). Player level

| | N | Mean | Median | SD | P25 | P75 |
|---|---|---|---|---|---|---|
| Move Quality | 1,241 | –2.20 | –2.18 | 0.67 | –2.52 | –1.79 |
| Number of Errors | 1,241 | 0.17 | 0.12 | 0.22 | 0.00 | 0.20 |
| Magnitude of the Critical Mistake | 1,241 | –6.20 | –5.98 | 2.09 | –6.96 | –5.05 |
| Age | 1,188 | 32.41 | 26.98 | 16.11 | 20.12 | 42.55 |
| Young | 1,188 | 0.50 | 1.00 | 0.50 | 0.00 | 1.00 |
| Rank | 1,104 | –0.52 | –0.52 | 0.31 | –0.79 | –0.26 |
| Rank Diff | 1,097 | 0.14 | 0.11 | 0.18 | 0.00 | 0.25 |

*Note*. This table provides descriptive statistics of variables at the player-game level in Panel (a) and at the player level in Panel (b). Note that, to ease the interpretation of results, we multiply negative one by the rank of a player and divide it by 1,000 (*Rank*). That is, a higher *Rank* value indicates a better player. We also divide the rank difference between the focal player and the opponent by 1,000 (*Rank Difference*). A negative value for *Rank Difference* indicates that the focal player is a better player.

**Table 2.** Effects of APG on average move quality of professional players: Time trend analysis

| Dependent Variable: | ***Move Quality*** | |
|---|---|---|
| Model: | (1) | (2) |
| *Variables* | | |
| Post | 0.756<br>(0.017)<br>[$p$<0.001] | –1.007<br>(0.038)<br>[$p$<0.001] |
| Trend | | 0.007<br>(0.003)<br>[$p$=0.023] |
| Post × Trend | | 0.116<br>(0.004)<br>[$p$<0.001] |
| *Fixed effects* | | |
| Player | Yes | Yes |
| Opponent Player | Yes | Yes |
| *Fit statistics* | | |
| Observations | 49,946 | 49,946 |
| $R^2$ | 0.264 | 0.330 |
| Within $R^2$ | 0.116 | 0.195 |

*Note.* This table shows the regression estimates on the effects of APG on the *Move Quality* of professional Go players before and after the first public release of an APG, Leela. *Post* takes unity for the games played in the quarters after February 2017. *Trend* refers to the number of quarters that had elapsed since the beginning of 2015; *Trend* takes the value of 10 in the first quarter after Leela's release (Q2 2017). Clustered standard errors at a focal-player level are in parentheses and p-values are in squared brackets.

**Table 3.** Differential effects of APG by player age:
Estimates on move quality and move match of young players compared to that of old players

| Dependent Variable: | *Move Quality* | | | | *Move Match* | | |
|---|---|---|---|---|---|---|---|
| Model: | (1) | (2) | (3) | (4) | (5) *k=1* | (6) *k=3* | (7) *k=5* |
| *Variables* | | | | | | | |
| Young | 0.096<br>(0.020)<br>[$p<0.001$] | –0.053<br>(0.021)<br>[$p=0.010$] | | | | | |
| Rank | 0.846<br>(0.036)<br>[$p<0.001$] | 0.828<br>(0.037)<br>[$p<0.001$] | 1.723<br>(0.246)<br>[$p<0.001$] | 2.582<br>(0.292)<br>[$p<0.001$] | 0.354<br>(0.048)<br>[$p<0.001$] | 0.318<br>(0.046)<br>[$p<0.001$] | 0.218<br>(0.037)<br>[$p<0.001$] |
| Rank Diff | 0.128<br>(0.028)<br>[$p<0.001$] | 0.120<br>(0.028)<br>[$p<0.001$] | 0.067<br>(0.025)<br>[$p=0.008$] | 1.040<br>(0.164)<br>[$p<0.001$] | 0.096<br>(0.027)<br>[$p<0.001$] | 0.075<br>(0.024)<br>[$p=0.002$] | 0.049<br>(0.022)<br>[$p=0.025$] |
| White | –0.133<br>(0.010)<br>[$p<0.001$] | –0.133<br>(0.010)<br>[$p<0.001$] | –0.130<br>(0.009)<br>[$p<0.001$] | –0.131<br>(0.009)<br>[$p<0.001$] | 0.045<br>(0.002)<br>[$p<0.001$] | 0.091<br>(0.002)<br>[$p<0.001$] | 0.092<br>(0.001)<br>[$p<0.001$] |
| 7.5 Komi | 0.021<br>(0.016)<br>[$p=0.191$] | 0.019<br>(0.016)<br>[$p=0.248$] | 0.021<br>(0.016)<br>[$p=0.190$] | 0.037<br>(0.018)<br>[$p=0.047$] | 0.001<br>(0.003)<br>[$p=0.680$] | 0.001<br>(0.003)<br>[$p=0.679$] | 0.000<br>(0.002)<br>[$p=0.940$] |
| Post × Young | | 0.268<br>(0.028)<br>[$p<0.001$] | 0.220<br>(0.031)<br>[$p<0.001$] | 0.203<br>(0.031)<br>[$p<0.001$] | 0.031<br>(0.005)<br>[$p<0.001$] | 0.025<br>(0.005)<br>[$p<0.001$] | 0.018<br>(0.004)<br>[$p<0.001$] |
| *Fixed effects* | | | | | | | |
| Quarter | Yes | Yes | Yes | Yes | Yes | Yes | Yes |
| Player | | | Yes | Yes | Yes | Yes | Yes |
| Opponent Player | | | | Yes | Yes | Yes | Yes |
| *Fit statistics* | | | | | | | |
| Observations | 47,292 | 47,292 | 47,292 | 47,292 | 47,292 | 47,292 | 47,292 |
| $R^2$ | 0.277 | 0.281 | 0.325 | 0.350 | 0.294 | 0.388 | 0.407 |
| Within $R^2$ | 0.065 | 0.070 | 0.013 | 0.014 | 0.031 | 0.120 | 0.157 |

*Note*. This table shows the regression estimates on the heterogeneous effects of APG by player age; the *Move Quality* of younger players compared to that of older players is estimated (Models 1–4). *Post* refers to games played in the quarters after the first public release of an APG in February 2017, and *Young* refers to young professional Go players. Models 5–7 assess the effect on *Move Match* by comparing players' moves with APG's top 1, 3, and 5 suggestions, respectively. Clustered standard errors at a focal-player level are in parentheses and p-values are in squared brackets.

**Table 4.** Effects of APG on move quality:
Errors and a critical mistake as mechanisms

| Dependent Variable: | ***Number of Errors*** | | ***Magnitude of the Critical Mistake*** | |
|---|---|---|---|---|
| Model: | (1) | (2) | (3) | (4) |
| *Variables* | | | | |
| Post | –0.055<br>(0.004)<br>[*p*<0.001] | 0.082<br>(0.013)<br>[*p*<0.001] | –1.430<br>(0.053)<br>[*p*<0.001] | 2.261<br>(0.143)<br>[*p*<0.001] |
| Trend | | 0.000<br>(0.001)<br>[*p*=0.730] | | –0.028<br>(0.012)<br>[*p*=0.021] |
| Post × Trend | | –0.009<br>(0.002)<br>[*p*<0.001] | | –0.233<br>(0.015)<br>[*p*<0.001] |
| *Fixed effects* | | | | |
| Player | Yes | Yes | Yes | Yes |
| Opponent Player | Yes | Yes | Yes | Yes |
| *Fit statistics* | | | | |
| Observations | 49,946 | 49,946 | 49,945 | 49,945 |
| $R^2$ | 0.077 | 0.081 | 0.123 | 0.145 |
| Within $R^2$ | 0.005 | 0.008 | 0.028 | 0.052 |

*Note*. This table shows the impact of APGs on errors and on the critical mistake by professional Go players before and after the release of Leela. A dependent variable for Models 1 and 2 is *Number of Errors* and for Models 3 and 4 is *Magnitude of the Critical Mistake*. *Post* refers to games played in the quarters after the first public introduction of the APG in February 2017, and *Trend* refers to the number of quarters passed since the first quarter in our sample. Clustered standard errors at a focal-player level are in parentheses and p-values are in squared brackets.

**Table 5.** Mediation analysis on game winning: Move quality, errors, and a critical mistake

| Dependent Variables: | ***Win*** | ***Move Quality*** | ***Number of Errors*** | ***Magnitude of the Critical Mistake*** | ***Win*** | | | | | | |
|---|---|---|---|---|---|---|---|---|---|---|---|
| Model: | (1) | (2) | (3) | (4) | (5) | (6) | (7) | (8) | (9) | (10) | (11) |
| *Variables* | | | | | | | | | | | |
| Rank | –0.095 | 2.582 | –0.350 | –7.373 | –0.096 | –0.027 | –0.066 | –0.161 | –0.110 | –0.140 | –0.167 |
| | (0.106) | (0.292) | (0.101) | (1.139) | (0.098) | (0.099) | (0.098) | (0.105) | (0.106) | (0.105) | (0.105) |
| | [*p*=0.372] | [*p*<0.001] | [*p*<0.001] | [*p*<0.001] | [*p*=0.327] | [*p*=0.782] | [*p*=0.499] | [*p*=0.125] | [*p*=0.300] | [*p*=0.183] | [*p*=0.111] |
| Rank Diff | –1.802 | 1.040 | 0.016 | –1.897 | –1.826 | –1.796 | –1.811 | –1.829 | –1.801 | –1.815 | –1.829 |
| | (0.087) | (0.164) | (0.065) | (0.692) | (0.086) | (0.086) | (0.086) | (0.087) | (0.087) | (0.087) | (0.087) |
| | [*p*<0.001] | [*p*<0.001] | [*p*=0.804] | [*p*=0.006] | [*p*<0.001] | [*p*<0.001] | [*p*<0.001] | [*p*<0.001] | [*p*<0.001] | [*p*<0.001] | [*p*<0.001] |
| White | 0.026 | –0.131 | 0.047 | 0.628 | 0.029 | 0.028 | 0.030 | 0.029 | 0.027 | 0.029 | 0.030 |
| | (0.005) | (0.009) | (0.004) | (0.037) | (0.005) | (0.005) | (0.005) | (0.005) | (0.005) | (0.005) | (0.005) |
| | [*p*<0.001] | [*p*<0.001] | [*p*<0.001] | [*p*<0.001] | [*p*<0.001] | [*p*<0.001] | [*p*<0.001] | [*p*<0.001] | [*p*<0.001] | [*p*<0.001] | [*p*<0.001] |
| 7.5 Komi | 0.000 | 0.037 | –0.013 | –0.129 | –0.001 | 0.000 | –0.001 | –0.001 | –0.001 | –0.001 | –0.001 |
| | (0.010) | (0.018) | (0.008) | (0.080) | (0.010) | (0.010) | (0.010) | (0.010) | (0.010) | (0.010) | (0.010) |
| | [*p*=0.990] | [*p*=0.047] | [*p*=0.093] | [*p*=0.107] | [*p*=0.945] | [*p*=0.978] | [*p*=0.958] | [*p*=0.918] | [*p*=0.949] | [*p*=0.930] | [*p*=0.906] |
| Post × Young | 0.024 | 0.203 | –0.024 | –0.487 | | | | 0.018 | 0.023 | 0.021 | 0.018 |
| | (0.009) | (0.031) | (0.009) | (0.103) | | | | (0.009) | (0.009) | (0.009) | (0.009) |
| | [*p*=0.008] | [*p*<0.001] | [*p*=0.006] | [*p*<0.001] | | | | [*p*=0.041] | [*p*=0.012] | [*p*=0.021] | [*p*=0.041] |
| Move Quality | | | | | 0.026 | | | 0.026 | | | 0.017 |
| | | | | | (0.002) | | | (0.002) | | | (0.003) |
| | | | | | [*p*<0.001] | | | [*p*<0.001] | | | [*p*<0.001] |
| Number of Errors | | | | | | –0.042 | | | –0.042 | | 0.003 |
| | | | | | | (0.006) | | | (0.006) | | (0.009) |
| | | | | | | [*p*<0.001] | | | [*p*<0.001] | | [*p*=0.695] |
| Magnitude of the Critical Mistake | | | | | | | –0.006 | | | –0.006 | –0.004 |
| | | | | | | | (0.001) | | | (0.001) | (0.001) |
| | | | | | | | [*p*<0.001] | | | [*p*<0.001] | [*p*<0.001] |
| *Fixed effects* | | | | | | | | | | | |
| Player | Yes | Yes | Yes | Yes | Yes | Yes | Yes | Yes | Yes | Yes | Yes |
| Quarter | Yes | Yes | Yes | Yes | Yes | Yes | Yes | Yes | Yes | Yes | Yes |
| Opponent Player | Yes | Yes | Yes | Yes | Yes | Yes | Yes | Yes | Yes | Yes | Yes |
| *Fit statistics* | | | | | | | | | | | |
| Observations | 47,280 | 47,292 | 47,292 | 47,291 | 47,538 | 47,538 | 47,537 | 47,280 | 47,280 | 47,279 | 47,279 |
| $R^2$ | 0.214 | 0.350 | 0.082 | 0.153 | 0.217 | 0.215 | 0.217 | 0.216 | 0.215 | 0.216 | 0.216 |
| Within $R^2$ | 0.018 | 0.014 | 0.005 | 0.011 | 0.020 | 0.019 | 0.020 | 0.020 | 0.019 | 0.020 | 0.021 |

*Note*. This table shows how *Move Quality* leads to winning a game. We test two mechanisms, *Number of Errors* and *Magnitude of the Most Critical Mistake*. Models 1 to 4, respectively, indicate that, after the release of the APG, younger professional Go players were more likely to win, to improve *Move Quality*, to decrease *Number of Errors*, and to reduce *Magnitude of the Critical Mistake*. A dependent variable for Models 5 through 11 is whether a player wins a game. Models 5 to 7, respectively, show a player is more likely to win a game if the player's *Move Quality* is greater, if the player's *Number of Errors* are fewer, and if the player has a smaller *Magnitude of the Most Critical Mistake*. The finding is robust when we account for the differences in *Move Quality* by age, as shown in Models 8 through 10. Model 11 presents the full specification that includes all relevant variables. Taken together, younger players improve *Move Quality*, decrease *Number of Errors*, and reduce *Magnitude of the Most Critical Mistake* after the introduction of the APG; these changes lead to eventually winning a game. Clustered standard errors at a focal-player level are in parentheses and p-values are in squared brackets